# Doppler Ghosts in a Hall of Mirrors:
# Polarisation Profiles of Scattered Emission Lines.

William J. Henney

Instituto de Astronomía, Universidad Nacional Autónoma de México

RESUMEN

Los corrimientos Doppler de las líneas de emisión ópticas que han sido dispersadas por el polvo y los electrones de los alrededores pueden dar información útil sobre la cinemática, geometría y condiciones físicas de los flujos astrofísicos. En teoría, los dispersores pueden proporcionar visiones desde diferentes direcciones del gas que está emitiendo líneas, así se puede determinar el campo de velocidades en tres dimensiones y observar fuentes que están escondidas y no pueden ser observadas directamente. Desafortunadamente, como en una sala de espejos, las imágenes múltiples resultantes pueden ser confusas y difíciles de interpretar. En general, la geometría de dispersión será desconocida, lo cual hace difícil separar el efecto del movimiento de los dispersores del efecto de movimiento de la fuente emisora. En esta situación, las observaciones espectropolarimétricas pueden ser de gran ayuda ya que la luz dispersada será polarizada parcialmente a un grado que depende del ángulo y de los detalles del proceso de dispersión. El análisis de los perfiles de polarización de las líneas de emisión dispersadas puede determinar entonces la orientación y la velocidad relativas de la fuente y los dispersores. Tales técnicas son aplicadas a líneas de emisión dispersadas en la región en frente de la cabeza del chorro estelar de HH 1, en donde se demuestra que las observaciones espectropolarimétricas admitirían una determinación independiente de la velocidad del choque de proa con respecto al material en frente de él, lo cual podría probar los modelos de erupciones múltiples del chorro estelar.

ABSTRACT

The Doppler shifts of optical emission lines which have been scattered by surrounding dust and electrons can provide useful information about the kinematics, geometry and physical conditions of astrophysical flows. In principle, the scatterers can provide views of the line-emitting gas from different directions, allowing the 3-d velocity field of the emitting gas to be determined and revealing sources which are hidden from direct view. Unfortunately, as in a Hall of Mirrors, the resultant multiple images can be confusing and hard to interpret. In general, the scattering geometry will be unknown, which makes it difficult to disentangle the effect of the motion of the scatterers from that of the motion of the emitting source. In this situation, spectropolarimetric observations can be a great help, since the scattered light will be partially polarised to a degree dependent on the angle of scattering and on the details of the scattering process. Analysis of the polarisation profiles of the scattered emission lines can then determine the relative orientation and velocity of the source and the scatterers. Such techniques are applied to the upstream scattered emission lines in the HH 1 jet, where it is shown that spectropolarimetric measurements would allow an independent determination of the bowshock velocity with respect to the upstream matter, which would test "multiple outburst" models of the jet.

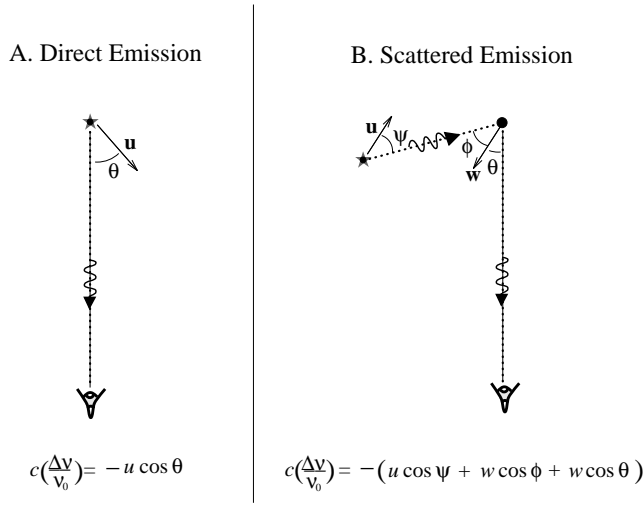

Fig. 1.— Comparison of the Doppler shifts seen in direct and in scattered emission lines.

## 1. INTRODUCTION: WHAT IS SPECIAL ABOUT SCATTERED EMISSION LINES?

Dust or electrons can act as mirrors, reflecting emission lines to provide multiple lines of sight to an emitting object. Apart from the opportunity to see objects that are hidden from direct sight, this has important consequences for the Doppler shifts of the scattered light. When an emission line is seen directly, the Doppler shift observed is a simple dot product of the source velocity with the unit vector toward the observer, whereas when the emission line is scattered, the resultant Doppler shift is a more complicated function of the velocities and relative positions of both the source and the scatterer (see Fig. 1).

**Problems of Interpretation:** Any of the following may be unknown:
 1. the geometry of the source(s) and scatterer(s),
 2. the velocity fields of the source(s) and scatterer(s),
 3. whether one is seeing scattering at all.

However, scattered light will be partially *polarised*, to a degree dependent on the scattering angle. Hence, by means of spectropolarimetry, scattered emission lines can be identified and useful kinematic information can be extracted.

Scattered optical emission lines have been detected in a wide variety of astrophysical objects, such as Herbig-Haro jets, in which light from a bowshock is reflected by upstream dust (Solf & Böhm 1991; Noriega-Crespo, Calvet & Böhm 1991; Henney, Raga & Axon 1994); the FU Orionis sytem Z CMa, in which the spectrum of a "hidden" companion is seen in polarised light (Whitney et al. 1993); the starburst galaxy M 82, in which a kiloparsec-scale reflection nebula scatters emission lines from the nuclear regions (Scarrott, Eaton & Axon 1991); and some Seyfert II galaxies, in which a hidden broad emission line region can be seen in polarised light that has been scattered by electrons that surround the nucleus (Miller & Goodrich 1990; Miller, Goodrich & Matthews 1991).

## 2. SOME SIMPLE KINEMATIC MODELS

To illustrate the effects that scattering has on line profiles, it is instructive to consider some simple kinematic models (for more details, see Henney 1994). In the first instance the scattering will be taken to be of the Rayleigh form (appropriate to small dust particles).

**Scattering Wind:** In this case, emission lines from a central source are scattered from a surrounding wind, outflowing at a constant speed. The scattered light is red-shifted only (with a maximum Doppler shift of twice the wind speed). This red wing is unpolarised if the wind is spherical.

**Moving Source:** In this case, a moving source emits lines which are scattered from a surrounding stationary dust cloud. The spatially integrated scattered light has red- and blue-shifted wings and the line is polarised, even for a spherically symmetric distribution of scatterers. The position angle of polarisation changes by 90° between the line core and the wings. The degree of polarisation is dependent on the inclination of the source velocity vector to the line of sight and is highest for a source moving in the plane of the sky.

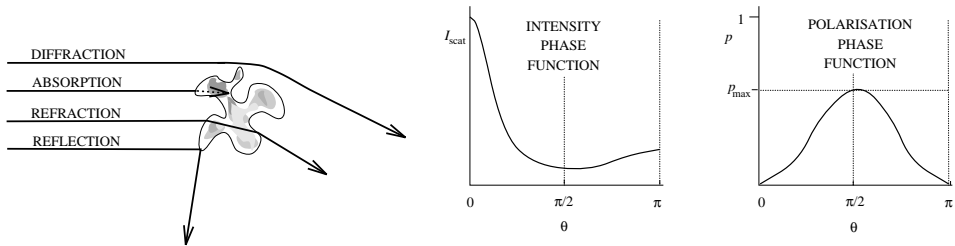

Fig. 2.— Scattering by a large dust particle. The right panels show the dependence on scattering angle of the intensity and polarisation of the scattered light for a typical "fluffy" particle of size comparable with the wavelength of the light (see West 1991; Giese et al. 1978).

**How do the physical properties of the scatterers affect the scattered emission lines?**

**Dust:** For scatterers small compared with the wavelength of the light, Rayleigh scattering occurs (front-back symmetrical, complete polarisation for 90° scattering) whereas with larger grains the scattering is much more complicated (see Fig. 2). There are two main differences from the Rayleigh case:

1. The diffracted component skews the scattered intensity distribution towards small scattering angles (diffraction fringes disappear when integrated over a size distribution or over the different orientations of an irregular scatterer).
2. The diffracted and refracted components are unpolarised (to first order), so the peak polarisation is reduced from unity.

For most purposes it is necessary to use empirical phase functions — the roughness and porosity of grains have a large effect on the angular scattering properties, especially the polarisation phase function (Kosaza et al. 1993; West 1991; Giese et al. 1978), and there is no computationally cheap way of calculating the phase functions of these "fluffy" grains in a physically realistic manner. The forward peaked phase function (see Fig. 2) tends to bias the scattered line profiles toward low Doppler shifts relative to the unscattered light. The scattered wings then have an intensity of roughly $I_s \sim \tau I_0 (1-g/1+g)^3$, where $g$ is the mean cosine of the scattering angle.

**Electrons:** High thermal speeds lead to very broad scattered line profiles. The electron thermal velocity distribution has a width $u_{\rm th} = \sqrt{4kT/m_{\rm e}} \simeq 800 \sqrt{T/10^4 {\rm K}}$ km s$^{-1}$. This will generally be greater than any bulk motion of the scattering medium and so the thermal broadening by the electrons cannot be ignored. The width of the broadening is dependent on the scattering angle. This is because, however fast the electrons are moving, they will not impart a large Doppler shift to the light if the scattering angle is small.

## 3. UPSTREAM DUST SCATTERING IN HH 1

Images of the head of HH 1 show what looks to be a bowshock and comparison with theoretical models (Raga, Barnes & Mateo 1990) suggest its motion is in the plane of the sky with a speed $v_{\rm BS} \sim 185$ km s$^{-1}$. However, proper motion measurements (Herbig & Jones 1981) indicate that the pattern speed of HH 1 is $v_{\rm PM} \sim 350$ km s$^{-1}$. The discrepancy between these two values can be understood if the bowshock travels into an environment which is already moving $v_{\rm env} \sim 150$ km s$^{-1}$. One possible mechanism for producing such a moving environment is a variable speed jet (Raga & Kofman 1992). Optical light from the region upstream of the bowshock was detected by Solf & Böhm (1991) and longslit spectroscopy indicates two components:

i. Narrow, slightly blue-shifted component. Visible in the lines H$\alpha$, [OI] and [OIII]. Probably intrinsic emission, possibly from a foreground nebula.
ii. Broad, very blue-shifted component ($v_{\rm shift} \sim 200$ km s$^{-1}$, FWHM$\sim 200$ km s$^{-1}$). Visible in the lines H$\alpha$, [OI] and [SII]. Probably light from the bowshock that has been scattered by upstream dust.

The work of Noriega-Crespo, Calvet & Böhm (1991) shows that a simple scattering model can plausibly explain the [SII] longslit spectrum. Here, the results of a detailed comparison between spectroscopic observations and a model in which emission lines from a moving hemispherical shell (an approximation to the bowshock) are scattered by an optically thin surrounding dust cloud) are briefly discussed .

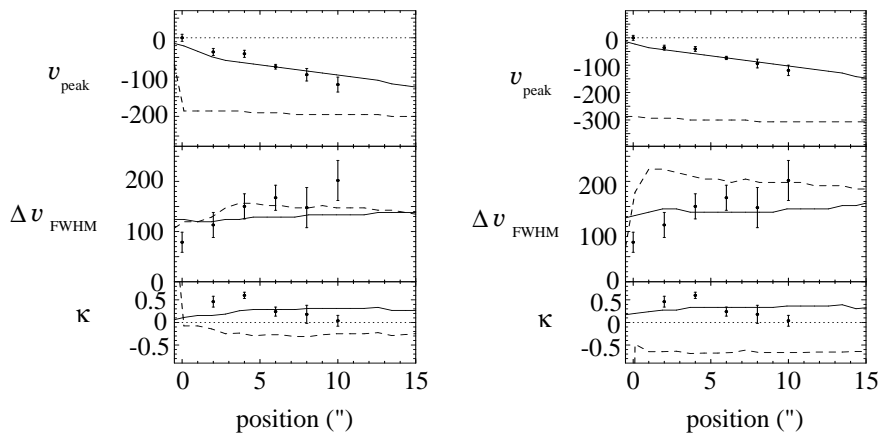

Fig. 3.— The velocity moments of the total scattered intensity (solid line) and the polarised intensity (dashed line) as a function of upstream distance from the source for two models that fit the data (filled circles) equally well. LEFT: $\alpha = 0°$, $v_{\rm BS} = 230\,\rm km\,s^{-1}$. RIGHT $\alpha = -6°$, $v_{\rm BS} = 330\,\rm km\,s^{-1}$. Note that spectropolarimetry would distinguish between these since they predict quite different values for the blue shift of the polarised intensity (dashed line, top panels).

**Application of Models:** Comparison of the source and scattered intensities suggest that the scattering optical depth of the source is $\tau_{\rm scat} \gtrsim 0.13$ (consistent with reddening measurements), which implies that $(R_c/15")(n_{\rm H}/750\,{\rm cm}^{-3}) \sim 3.5$, where $R_c$ and $n_{\rm H}$ are the angular radius and the hydrogen number density of the cloud. Then, comparison of various models with the intensity gradient of the scattered light along the slit of the [SII] spectrogram shows that *any* type of scattering phase function or shape of cloud can fit the data by adjusting the assumed cloud radius. However, when the models are compared with the spectral shapes of the scattered lines (from the longslit spectrograms, the velocity shift FWHM and skewness of the lines can be estimated as a function of position along the slit), the only acceptable fits are obtained with a model with a rather forward-peaked scattering phase function ($g = 0.66$), see Fig. 3. The speed of the emitting source (gas cooling behind the bowshock) is still not determined uniquely since it is dependent on the angle $\alpha$ between the source velocity vector and the plane of the sky. Two possibilities are: (i) $\alpha = 0°$, $v_{\rm BS} = 230\,\rm km\,s^{-1}$; (ii) $\alpha = -6°$, $v_{\rm BS} = 330\,\rm km\,s^{-1}$. The first of these lends some support to the "moving medium hypothesis", the second none at all. However, spectropolarimetry of the scattered lines would unambiguously distinguish between the two cases since the peak of the *polarised* intensity is a robust indicator of the source speed (Fig. 3).

If (i) is correct $\Rightarrow$ upstream environment is moving (multiple outbursts?)

If (ii) is correct $\Rightarrow$ bowshock models are wrong!

These results are described in more detail in Henney, Raga & Axon (1994).

REFERENCES

Giese, R. H., Weiss, K., Zerull, R. H., & Ono, T. 1978, A&A, 65, 265
Henney, W. J. 1994, ApJ, in press
Henney, W. J., Raga, A. C., & Axon, D. J. 1994, ApJ, in press
Herbig, G. H., & Jones, B. F. 1981, AJ, 86, 1232
Kosaza, T., Blum, J., Okamoto, H., & Mukai, T. 1993 A&A, 276, 278
Miller, J. S., & Goodrich, R. W. 1990, ApJ, 355, 456
Miller, J. S., Goodrich, R. W., & Matthews, W. G. 1991, ApJ, 378, 47
Noriega-Crespo, A., Calvet, N., & Böhm, K.-H. 1991, ApJ, 379, 676
Raga, A. C., Barnes, P. J., & Mateo, M. 1990, AJ, 99, 1919
Raga, A. C., & Kofman, L. 1992, ApJ, 386, 222
Scarrott, S. M., Eaton, N., & Axon, D. J. 1991, MNRAS, 252, 12P
Solf, J., & Böhm, K. H. 1991, ApJ, 375, 618
West, R. A. 1991, App. Optics, 30, 5316
Whitney, B. A., Clayton, G. C., Schulte-Ladbeck, R. E., Calvet, N., Hartmann, L., & Kenyon, S. J. 1993 ApJ, 417, 687